\documentclass[runningheads]{llncs}

\usepackage[hidelinks]{hyperref}
\usepackage{graphicx}
\usepackage{cite}

\usepackage{enumitem}

\usepackage{ amsmath }

\usepackage{ amsfonts }

\usepackage{ stmaryrd }

\usepackage{float}

\newcommand{\bm}[1]{\mathbf{#1}}

\usepackage{tikz}

\usepackage{amssymb}

\usetikzlibrary[decorations.pathreplacing]

\begin{document}
\title{NP Decision Procedure for Monomial and Linear Arithmetic Constraints}
\titlerunning{NP Decision Procedure for Monomial and Linear Arithmetic Constraints}
%
\author{Rodrigo Raya\orcidID{0000-0002-0866-9257} \and Jad Hamza \and  \\
Viktor Kun\v{c}ak\orcidID{0000-0001-7044-9522}}
%
%
\institute{School of Computer and Communication Science \\ École Polytechnique Fédérale de Lausanne (EPFL), Switzerland 
\email{\{rodrigo.raya,viktor.kuncak\}@epfl.ch}, \email{jad.hamza@protonmail.com}}
\maketitle              
\sloppy
\begin{abstract}
Motivated by satisfiability of constraints with function symbols, we consider numerical inequalities on non-negative integers. The constraints we consider are a conjunction of a linear system $Ax=b$ and a conjunction of (non-)convex constraints of the form $x_i \ge x_j^n$ ($x_i \le x_j^n$).
We show that the satisfiability of these constraints is NP-complete even if the solution to the linear part is given explicitly. As a consequence, we obtain NP completeness for an extension of certain quantifier-free constraints on sets with cardinalities (QFBAPA) with function images $S=f[P^n]$.
\end{abstract}


\section{Introduction}

Many satisfiability problems in logic naturally reduce to
numerical constraints. This includes in particular
two-variable logic with counting \cite{DBLP:conf/csl/Pratt-Hartmann14, DBLP:journals/mlq/Pratt-Hartmann15, DBLP:journals/jolli/Pratt-Hartmann05}, as well as description logics with cardinality bounds \cite{10.1145/3297280.3297390,DBLP:conf/dlog/BaaderBR19}. In many of these cases, the resulting numerical constraints belong to linear integer arithmetic (LIA). However, satisfiability in the presence of functions with multiple arguments naturally leads to multiplicative constraints \cite{givan_tarskian_2002,yessenov_collections_2010}. Perhaps due to a negative answer to Hilbert's 10th problem, such multiplicative constraints are often avoided. We show, however, that certain classes of such constraints can still be solved within the complexity class NP---arguably low complexity for logical constraints.

\noindent \textbf{Prequadratic constraints.} The main class of numerical constraints we consider extends LIA with atoms of the form $x \le y^n$. It is a strict subset of the so-called prequadratic constraints, which also allow atoms of the form $x \le yz$ and were first studied in \cite{givan_tarskian_2002}. Two decades ago, the authors of \cite{givan_tarskian_2002} sketched an argument that prequadratic constraints can be decided in NEXPTIME and conjectured that the complexity can be reduced to NP. However, no result showing membership in NP has appeared to date. In the meantime, an alternative method was used to settle the complexity for Tarskian constraints~\cite{mielniczuk_tarskian_1998}. Nevertheless, other reductions to such non-linear inequalities remain of interest.

In \cite{DBLP:conf/rta/KobayashiO08}, the authors prove the decidability of 
satisfiability of \emph{monotone exponential Diophantine formulas}
(LIA with atoms of the form $x \leq y^z$ and of the form $x \leq yz$).
They do so by reducing it to the emptiness of \emph{monotone AC-tree automata}
(tree automata modulo associativity and commutativity), but they do not
provide complexity bounds.

One application of non-linear inequalities is the satisfiability of set algebra with cardinality constraints and images of functions of multiple arguments \cite{yessenov_collections_2010}, which is related to description logics \cite{DBLP:conf/dlog/2003handbook}. Consider the constraint $A = f[B,C]$, which states that $A$ is the image of a two-argument function $f$
under sets $B$ and $C$. Assume that all sets are non-empty. Then such $f$ exists if and only if $|A| \le |B| |C|$, where the equality is reached only when $f$ is injective. Denoting $|A|$ by $x$, $|B|$ by $y$ and $|C|$ by $z$, we obtain constraints of the form $x \leq y z$. What is more, by picking fresh sets $A$, $B$, $C$, we can express arbitrary conjunctions of such constraints. In other words, solving numerical inequalities is necessary to check certain constraints of cardinalities and function images.

While we leave open the question of NP membership for the general case, $x \le yz$, we solve it in the case of conjunctions of constraints of the form $x \le y^2$, and, more generally, $x \le y^n$ for any positive integer $n$.
We also consider the dual case, $x \ge y^2$, and more generally, $x \ge y^n$.
As an application, we describe logics that handle quantifier-free constraints on sets with cardinalities (QFBAPA) and (inverse) function images $S=f[P^n]$ ($S = f^{-1}[P^n]$). The atomic formula $S=f[P^n]$ ($S = f^{-1}[P^n]$) expresses that $S$ is an (inverse) image of $P^n$ under function $f$. As a consequence of the results shown for $x \le y^n$ ($x \ge y^n$), under restrictions on multiple occurrences of $f$, the satisfiability problem of these logics with (inverse) function images is in NP.

We believe that such results are of interest because they compose with other constructions that preserve NP membership. In particular, in a recent analysis of array theories \cite{raya_np_2022} we observed that the fragment of combinatory array logic \cite{de_moura_generalized_2009} corresponds to the theory generated by a power structure with an arbitrary index set subject to QFBAPA constraints. Given that \cite{raya_np_2022} shows a NP complexity bound for such product, it is natural to ask how far we can extend NP satisfiability results. The non-linear constraints we present in this paper can be applied to the case when the index set $I$ is a power $J^n$, because image constraints with functions on subsets of $J^n$ reduce to non-linear constraints whose complexity we consider.

Finally, we argue that non-linear inequalities are such a natural and fundamental problem that their complexity is of intrinsic interest. Once their complexity is understood, they are likely to find other applications.

\noindent \textbf{(Non-)convexity.}
\cite{tarasov_bounds_1980} has proven an NP complexity bound for certain classes of convex non-linear constraints. However, the class of numerical constraints considered in our Theorem~\ref{thm:more} is different, since we do not bound the degree of the non-linear monomial. On the other hand, the class of numerical constraints considered in Theorem~\ref{thm:less} is non-convex. Indeed, consider the constraint $x \le y^2$. Both $(x,y) = (4,2)$ and $(x,y) = (16,4)$ satisfy the constraints, but the midpoint of the line segment connecting them is $(10,3)$, which does not satisfy the constraint. We are not aware of previous NP complexity bounds for such non-convex constraints.

\noindent \textbf{Extending linear constraints while retaining NP complexity.}
It has been long known that the satisfiability problem of quantifier-free linear integer arithmetic constraints is
NP-complete \cite{gathen_bound_1978, kannan_computational_1978, papadimitriou_complexity_1981}. 

If we add multiplication with atoms of the form 
$t = t_1 t_2$, the problem becomes undecidable because it expresses general multiplication
and thus Hilbert's 10th problem~\cite{yuri_matijasevich_enumerable_1970}.
Moreover, even the case of atoms $t' = t^2$ yields
undecidability~\cite[Section~2.3]{DBLP:conf/stacs/BursucCD07}, because of 
the identity $2t_1t_2 = (t_1 + t_2)^2 - t_1^2 - t_2^2$.  
It is thus natural to explore the limits of the decidability of quadratic 
constraints. 

\noindent \textbf{Organization of the paper.} Section~\ref{section:background} introduces the classes of constraints that we solve. They are of the form $\varphi = L \land Q$ where $L$ stands for linear constraints and $Q$ for certain conjunctions of monomial inequalities. We also recall known facts on the structure of semilinear sets. Finally, Lemma~\ref{lem:normal} gives a normal form that is used in the rest of the paper. Section~\ref{section:more} proves a
NP complexity bound when $Q$ is a conjunction of constraints of the form $x \geq y^n$. Section~\ref{section:less} proves a NP complexity bound when $Q$ is a conjunction of 
atoms of the form $x \leq y^n$. Note that one cannot reduce neither case to the other because non-negativity of numbers breaks the symmetry between $\le$ and $\ge$ (in fact,
one case has a small model property whereas the other one needs certificates that
are not always actual values of integer variables).
Section~\ref{section:hardness} proves NP-hardness of both problems even under the assumption that the solution of the linear part is given explicitly. Section~\ref{section:logical} gives the complexity of satisfiability for sets with cardinalities and (inverse) function images based on Sections~\ref{section:more} and \ref{section:less}. Section~\ref{section:conclusion} concludes the paper.


\section{Background and Initial Analysis}
\label{section:background}

\subsection{Basic definitions and facts} \label{section:semilinear}

\textbf{Families of linear arithmetic constraints.} We now give precise definitions of the families of constraints that we discuss in the paper. In the following, $\mathbb{N}$ will denote the set of non-negative integer numbers. Our constraints do not use quantifiers and can be fully expressed in the framework of relational logic \cite[Chapter 4]{burris_logic_1997}, that is, first-order logic without quantifiers. All the families of constraints we address extend linear arithmetic, a restriction of full arithmetic that omits multiplication.

\begin{definition}
A linear arithmetic formula is a relational formula whose atoms are of the form $a_1 x_1 + \ldots + a_n x_n \le b$ where $a_1,\ldots,a_n$ and $b$ are integer constants and $x_1, \ldots, x_n$ are non-negative integer variables. 
\end{definition}

Note that we choose our variables over the non-negative integers since they represent cardinalities of sets. It is straightforward to reduce linear arithmetic constraints over the integers to those over non-negative integers by encoding each integer variable as the difference of two non-negative integer variables. As we mentionned, the satisfiability problem of linear arithmetic constraints is in NP. In this paper, we will show NP-completeness for the following extensions of linear arithmetic.

\begin{definition}
A less-than-monomial (more-than-monomial) constraint is a relational conjunction whose atoms are linear arithmetic formulae or of the form $x \le y^n$ ($x \ge y^n$) where $x,y$ are variables, $n$ is a non-negative integer that may be distinct for different atoms and $y^n$ denotes the $n^{\textit{th}}$ power of $y$. 
\end{definition}

We will refer to the non-linear part of less-than-monomial and more-than-monomial constraints as \textit{monomial inequalities}. The non-linear restrictions of less-than-monomial constraints form a strict subset of the non-linearities in the prequadratic class \cite{givan_tarskian_2002}.

\begin{definition}
A set of Diophantine inequalities of the form $p(x_1,\ldots,x_n) \le q(x_1,\ldots,x_n)$ between polynomials $p$ and $q$ over nonnegative integer variables $x_1,\ldots,x_n$ is prequadratic if every $p$ is linear and every $q$ is either linear or is a product of variables.
\end{definition}

By adding slack variables, we may transform any prequadratic constraint $p(x_1,\ldots,x_n) \le q(x_1,\ldots,x_n)$ as a Diophantine equation $p(x_1,\ldots,x_n) + s = q(x_1,\ldots,x_n)$. Solving these equations over the integers was shown to be undecidable in a joint effort of Davis, Matiyasevich, Putnam and Robinson \cite{hilbert10}, which yielded a solution of the so-called Hilbert's tenth problem. Furthermore, it is easy to show  that the analogous problem over the non-negative numbers is also undecidable \cite[Section~1.3]{hilbert10}. In our case, this yields at once the following corollary.

\begin{corollary}
The satisfiability problem for relational conjunctions whose atoms are linear arithmetic formulae or of the form $x \le y^n$, $x \ge y^n$ where $x,y$ are variables, $n$ is a non-negative integer that may be distinct for different atoms and $y^n$ denotes the $n^{\textit{th}}$ power of $y$ is undecidable. 
\end{corollary}

In Section \ref{section:logical}, we will also make use of the quantifier-free fragment of BAPA \cite{kuncak_deciding_2006, kuncak_towards_2007}, termed QFBAPA, whose language allows to express Boolean algebra and cardinality constraints on sets. Figure \ref{fig:qfbapa-syntax} shows the syntax of the fragment: $F$ presents the Boolean structure of the formula, $A$ stands for the top-level constraints, $B$ gives the Boolean restrictions and $T$ the Presburger arithmetic terms. $\mathcal{U}$ stands for the universe of the interpretation and $\text{MAXC}$ for its cardinality.

\begin{figure}[H]
\centering
\begin{align*}
F & ::= A \, | \, F_1 \land F_2 \, | \, F_1 \lor F_2 \, | \, \lnot F \\
A & ::= B_1 = B_2 \, | \, B_1 \subseteq B_2 \, | \, T_1 = T_2 \, | \, T_1 \le T_2 \, | \, K \text{ dvd } T \\
B & ::= x \, | \, \emptyset \, | \, \mathcal{U} \, | \, B_1 \cup B_2 \, | \, B_1 \cap B_2 \, | \, B^c \\
T & ::= k \, | \, K \, | \, \text{MAXC} \, | \, T_1 + T_2 \, | \, K \cdot T \, |  \, |B| \\
K & ::= \ldots \, | \, -2 \, | \, -1 \, | \, 0 \, | \, 1 \, | \, 2 \, | \, \ldots
\end{align*}
\caption{QFBAPA's syntax}
\label{fig:qfbapa-syntax}
\end{figure}

Note that QFBAPA constraints can also be seen as extending linear arithmetic restrictions. Indeed, as noted in \cite[section 2]{kuncak_deciding_2006}, the addition of the cardinality operator allows to express all Presburger arithmetic (i.e. the theory of the structure $\langle \mathbb{N}, 0, 1, +, \le \rangle$) operations. In turn, these can be efficiently represented by linear arithmetic constraints. The relation $K \text{ dvd } T$ (divisibility by an integer constant $K$) and the term $K \cdot T$ (multiplication by an integer constant) are added to preserve the expressive power of full first-order Presburger arithmetic as in \cite{presburger_uber_1929}.

\noindent \textbf{Semilinear sets.} As a first step of an NP procedure, we will guess a normal form of the input constraint. The particular normal form is based on results on the structure of the sets defined by the linear part of the constraint. 

Let $\mathbb{N}^n$ denote the direct product $\mathbb{N}$ taken $n$ times. We will distinguish elements of $\mathbb{N}^n$ from those in $\mathbb{N}$ using bold font. If $\bm{x} \in \mathbb{N}^n$ then the sum norm and the infinite norm are defined as follows:

\begin{align*}
\|\bm{x}\|_1 &= \sum_{i = 1}^n |x_i| \\
\|\bm{x}\|_{\infty} &= \max\{|x_1|,\ldots,|x_n|\}
\end{align*}

A subset $L \subseteq \mathbb{N}^n$ is \emph{linear} if there exist members $\bm{a}, \bm{b}^1, \ldots, \bm{b}^m \in \mathbb{N}^n$ such that:
\[
L = \Big\{\bm{x} \Big| \exists \alpha_1,\dots,\alpha_m \in \mathbb{N}.\ 
\bm{x} = \bm{a} + \sum_{i = 1}^m \alpha_i \bm{b}^i \Big\}
\]
The element $\bm{a}$ is called the \emph{base vector} of $L$ and the elements $\bm{b}^1, \ldots, \bm{b}^m$ are called the \emph{step vectors} of $L$. We refer to both the base vectors and the step vectors as the \emph{generators} of the set $L$. 

$S$ is \emph{semilinear} \cite[definition 12]{parikh_language_1961} if it is the union of a finite number of linear sets. The base vectors (step vectors, generators) of $S$ are defined as the union of the set of base vectors (step vectors, generators) of each of its linear parts \cite{piskac_linear_2008}. 

In \cite{ginsburg_semigroups_1966}, it was shown that the sets definable by linear arithmetic formulas are precisely the semilinear sets. Every semilinear set can be written in the form $\{\bm{x} \in \mathbb{N}^n | F(\bm{x}) \}$ where $F$ is a linear arithmetic formula. Furthermore, it was shown in \cite{loic_pottier_minimal_1991,tarlecki_solving_1991} that when given in terms of a linear arithmetic formula $F$, the semilinear set  defined by $F$ can be represented using a set of generators whose coefficients are polynomially bounded in the size of $F$. \cite[Theorem 2.13]{piskac_decision_2011} derives the following normal form for $F$ based on these facts.

\begin{theorem} \label{thm:semnf}
Let $F$ be a linear arithmetic formula of size $s$. Then there exist numbers $m, q_1, \ldots, q_m \in \mathbb{N}$ and vectors $\bm{a_i}, \bm{b_{ij}} \in \mathbb{N}^n$ for $1 \le j \le q_i, 1 \le i \le m$ with $\|\bm{a_i}\|_1, \|\bm{b_{ij}}\|_1 \le 2^{p(s)}$ with $p$  polynomial such that $F$ is equivalent to the formula:
\[
\exists \alpha_{11}, \ldots, \alpha_{mq_m}. \bigvee_{i = 1}^m \Big(\bm{x} = \bm{a_i} + \sum_{j =  1}^{q_i} \alpha_{ij} \bm{b_{ij}}\Big)
\]
\end{theorem}

Finally, the integer analog of Carathéodory's theorem \cite{eisenbrand_caratheodory_2006} allows to express any element of a semilinear set using polynomially many step vectors. It is formulated in terms of integer conic hulls: 

\begin{definition}
Given a subset $S \subseteq \mathbb{R}^n$, the integer conic hull of $S$ is the set: 
\[
int_{cone}(S) = \Bigg\{ \sum_{i = 1}^t \lambda_i s_i \Big| t \ge 0, x_i \in S, \lambda_i \in \mathbb{N} \Bigg\}
\]
\end{definition}

\begin{theorem}
Let $X \subseteq \mathbb{Z}^n$ be a finite set of integer vectors
and $\bm{b} \in int_{cone}(X)$. Then there exists a subset $X' \subseteq X$ such that $\bm{b} \in int_{cone}(X')$ and:
\[
|X'|\le 2n\log(4nM)
\]
where $M = \max_{\mathbf{x} \in X} \|\mathbf{x}\|_{\infty}$.
\end{theorem}

\noindent \textbf{Computational complexity.} We assume basic definitions in the theory of computation \cite{arora_computational_2009, sipser_introduction_2012}. We will use the notion of polynomial-time verifier which is equivalent to that of non-deterministic polynomial-time procedure:

\begin{definition}
A language $L \subseteq \{0,1\}^*$ is in NP if there exists a polynomial $p: \mathbb{N} \to \mathbb{N}$ and a polynomial-time Turing machine $M$, called the verifier for $L$ such that for every $x \in \{0,1\}^*$, $x \in L$ if and only if there exists $u \in \{0,1\}^{p(|x|)}$ such that $M(x,u) = 1$. If $x \in L$ and $u \in \{0,1\}^{p(|x|)}$ satisfy $M(x,u) = 1$, then $u$ is called a certificate for $x$.
\end{definition}

We will also use the definition of NP-hardness and NP-completeness. The notion of NP-reduction will be used to combine the normal form of Lemma~\ref{lem:normal} with the polynomial-time verifiers of Theorem~\ref{thm:more} and Theorem~\ref{thm:less} into a single NP procedure:

\begin{definition}
A language $L \subseteq \{0,1\}^*$ is NP-reducible to a language $L' \subseteq \{0,1\}^*$, if there is a non-deterministic polynomial-time computable function $f: \{0,1\}^* \to \{0,1\}^*$ such that for every $x \in \{0,1\}^*$, $x \in L$ if and only if $f(x) \in L'$.
\end{definition}

\subsection{Normal Form of Constraints}
\label{section:normal}

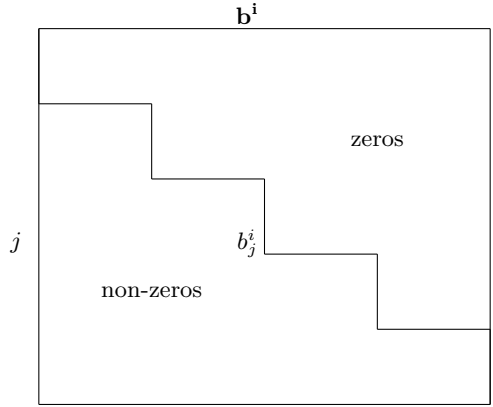
\begin{figure}[!ht]
\centering

\begin{tikzpicture}[xscale=1.5]
\node at (3,-1.5) { zeros };
\node at (1,-3.5) { non-zeros };

\node at (-0.2,-0.1) {  };
\node at (-0.2,-0.85) {  };
\node at (-0.2,-1.85) {  };
\node at (-0.2,-2.85) { $j$ };
\node at (-0.2,-3.85) {  };
\node at (-0.2,-4.85) {  };

\node at (0.15,0.2) {  };
\node at (0.85,0.2) { };
\node at (1.85,0.2) { $\bm{b^{i}}$ };

   \node at (1.85,-2.85) { $b^i_j$ };

\node at (3.85,0.2) { };

\draw (0,0) -- (0,-5);
\draw (0,-5) -- (4,-5);
\draw (4,-5) -- (4,0);
\draw (0,0) -- (4,0);
\draw (0,0) -- (0,-1);
\draw (0,-1) -- (1,-1);
\draw (1,-1) -- (1,-2);
\draw (1,-2) -- (2,-2);
\draw (2,-2) -- (2,-3);
\draw (2,-3) -- (3,-3);
\draw (3,-3) -- (3,-4);
\draw (3,-4) -- (4,-4);
\draw (4,-4) -- (4,-5);
\end{tikzpicture}

\caption{Vertical column arrangement of the step vectors $\bm{b}^1$,\dots,$\bm{b}^K$.}
\label{fig:stairs}
\end{figure}

The following lemma gives a normal form for linear arithmetic formulae conjoined with monomial inequality constraints. The resulting structure of the generators of the semilinear set is shown in Figure~\ref{fig:stairs}. This normal form will be used as input in Sections~\ref{section:more} and \ref{section:less} to establish NP complexity bounds for the more-than-monomial and less-than-monomial constraints. 


\begin{lemma} \label{lem:normal}
There is an NP reduction mapping each satisfiable formula $(F \land Q)(\bm{x})$ where $F$ is a linear arithmetic formula and $Q$ is a conjunction of monomial inequalities to a satisfiable formula $(L \land Q')(\bm{x})$ where $L$ is of the form $\exists \alpha_1, \ldots, \alpha_K \in \mathbb{N}. \bm{x} = \bm{a} + \sum_{i = 1}^K \alpha_i \bm{b}^i$ and $Q'$ is a conjunction of monomial inequalities obtained from $Q$ permuting its variables. Moreover, the reduction ensures that:

\begin{enumerate}
    \item $K$ is polynomial in the size of $F$.
    \item If $x_i,a_i$ and $b^i_j$ are the coordinates of the vectors $\bm{x}, \bm{a}$ and $\bm{b}^i$ then:
    
    \begin{itemize}
        \item $a_1 \le \ldots \le a_n$ and $b_1^i \le \ldots \le b_n^i$ for all $i = 1, \ldots, K$.
        \item for all satisfying assignments of $L \land Q'$, $x_1 \le \ldots \le x_n$.
    \end{itemize}
    
    \item $\bm{b}^i > \bm{b}^{i+1}$ for all $i = 1,\ldots,K$ where $>$ is the strict lexicographic order defined as $\bm{b} > \bm{b}' \equiv \exists k.\ 
      b_k > b'_k \land \bigwedge_{1 \le j < k} b_j = b'_j$.
    \item $b_1^i = 0$ for all $i = 1,\ldots, K$.
\end{enumerate}
\end{lemma}
\begin{proof}
We give an NP reduction composing the following two transformations.

The first transformation ensures that the components of a solution $\bm{x}$ are linearly ordered. For this purpose, we observe that the formula $(F \land Q)(\bm{x})$ is satisfiable if and only if there exists a permutation $\sigma$ such that:
\[
(F \land Q)(\bm{x}) \land x_{\sigma(1)} \leq \dots \leq x_{\sigma(n)}
\]
is satisfiable. An polynomial-time verifier can encode $\sigma$ in its certificate and compute
$(F \land Q)(\sigma^{-1}(\bm{y})) \land y_1 \le \ldots \le y_{n}$ where $\sigma^{-1}(\bm{y}) = (y_{\sigma^{-1}(1)}, \ldots, y_{\sigma^{-1}(n)})$. This results in a new system $(F' \land Q')(\bm{y})$ of linear and monomial inequalities which we use as input in the subsequent step.

The second transformation computes the generators of the semilinear set $S'$ defined by the formula $F'$. For this we use the facts we recalled in Section~\ref{section:semilinear}. Note that the components of the generators are computable by a polynomial-time verifier by Theorem \ref{thm:semnf}. Since $S' = \cup_{i = 1}^n L_i$ where each $L_i$ is a linear set, it follows that a solution must lie in some of the linear components $L_{i_0}$. Thus, the certificate of the polynomial-time verifier may only include the representation of $L_{i_0}$. By \cite{eisenbrand_caratheodory_2006}, if $\bm{y_0} \in L_{i_0}$ is a solution of $(F' \land Q')(\bm{y})$ then we may find a linear subset $L \subseteq L_{i_0}$ requiring only polynomially many generators such that $\bm{y_0} \in L$. If there is no solution of $(F' \land Q')(\bm{y})$ it is clear that no such subset can contain one. Thus, we can guess the generators of $L$ with a polynomial-time verifier.

Composing the two transformations, we obtain a formula:
\[
\psi \equiv \Bigg(\exists \alpha_1, \ldots, \alpha_K \in \mathbb{N}. \bm{y} = \bm{a} + \sum_{i = 1}^K \alpha_i \bm{b}^i\Bigg) \land Q'(\bm{y})
\]
equisatisfiable with $(F \land Q)(\bm{x})$ and with $K$ polynomial in the size of $F$.

We now show items 2, 3 and 4. 

For 2), observe that the first transformation ensures that $y_1 \le \ldots \le y_n$ for any solution $\bm{y} = \bm{a} + \sum_{i = 1}^K \alpha_i \bm{b}^i$. Taking all the $\alpha_i = 0$ yields $\bm{y} = \bm{a}$. This implies that $a_1 \le \ldots \le a_n$. Now, for each $i$, take $\bm{y} = \bm{a} + \alpha_i \bm{b}^i$ by setting $\alpha_j = 0$ for $j \neq i$. Finally, the coordinates of $\bm{b}^i$ are increasing. By contradiction, if $b_j^i > b_k^i$ for $j < k$ then there is some $\alpha_i$ such that $y_j = a_j + \alpha_i b_j^i > a_k + \alpha_i b_k^i = y_k$. But we showed that the components of $\bm{y}$ are linearly ordered.

For 3), observe that there is no need to have two identical (or even linear dependent) vectors among $\bm{b}^i$ in $\psi$. So we assume the vectors are distinct. As the order of vectors does not matter either, we will henceforth assume that the order of vectors is chosen so that $i_1 < i_2$ implies $\bm{b}^{i_1} > \bm{b}^{i_2}$, i.e. $\bm{b}^1 > \ldots > \bm{b}^K$. 

4) is a consequence of the coefficients of the step vectors being linearly ordered.
Indeed, if for some $i$ we have that $b_1^i \ge 1$ then, for all $j$, $b^i_j \ge b^i_1 \ge 1$.
Setting $\alpha_j = 0$ for $j \neq i$ and letting $\alpha_i$ increase towards infinity, each prequadratic
constraint $x_i \le x_j x_k$ becomes satisfied because the left-hand side grows linearly whereas the right-hand side grows
quadratically.  
This implies that $x_1 = a_1$.
\end{proof}

Figure~\ref{fig:stairs} shows the structure of the matrix of step vectors that the verifier of Lemma~\ref{lem:normal} guesses. The matrix is syntactically similar to the row echelon form found in Gaussian elimination. Here, all zero rows are at the top and each zero value of a row appears to the right (but not necessarily strictly to the right) of its previous row.

\section{Satisfiability of Convex Monomials}
\label{section:more}

This section proves an NP bound for more-than-monomial constraints. We assume the input formula is given in the normal form found in Lemma~\ref{lem:normal}. Let $m$ denote the largest constant appearing in the constraint, that is, the largest among all coordinates $a_j$ and $b^i_j$. We first note that, in this case, it is not possible to find a polynomial bound on the size of minimal solutions, because there are systems whose minimal solutions are doubly exponential in $m$ (and thus have an exponential number of bits). For example, consider the following system of $n$ variables:
\[
\begin{cases}
x_1 \ge 2 \\
x_{i+1} \ge x_i^2 & \forall i \in \{1, \ldots, n-1\} \\
\end{cases}
\]
Consider any solution $x_1,\ldots,x_n$ of the above system. Then by induction it immediately follows that
$x_i \ge 2^{2^{i-1}}$ for $1 \le i \le n$. Indeed, $x_1 \ge 2 = 2^{2^0}$ and if $x_i \ge 2^{2^{i-1}}$ for $i < n$ then:
\[
    x_{i+1} \ge x_i^2 \ge \left( 2^{2^{i-1}} \right)^2 = 2^{2 \cdot 2^{i-1}} = 2^{2^i} 
\]
Despite the lack of small enough solutions, we show that the satisfiability problem can be solved in non-deterministic polynomial time by observing that satisfiability can be checked without exhibiting a specific solution.
In Section~\ref{section:logical}, we apply this result to show an NP upper bound of a fragment of quantifier-free constraints on sets with cardinalities (QFBAPA) and inverse function images which we term QFBAPA-InvFun.

\begin{theorem} \label{thm:more}
Satisfiability of more-than-monomial constraints is in NP.
\end{theorem}
\begin{proof}
We can assume that the input formula is of the form specified in Lemma~\ref{lem:normal}. Let $m$ denote the maximum of the coefficients of the generators of the linear part.
We introduce the notation $j^*$ to refer to the column of the first zero entry for the $j$-th row, and $supp(j)$ to refer to the set of indices with non-zero values of the $j$-th row of the step vector matrix (see Figure \ref{fig:stairswithsupport}): 
\begin{align*}
j^* &:= 
\begin{cases}
0 & \text{ if for every } 1 \le i \le k.\ b_j^i \neq 0 \\
i & \text{ if $i$ is the least index such that } b_j^i = 0
\end{cases} \\
\text{supp}(j) &:= \{i \mid b_j^i \neq 0 \} = [1,j^*-1]
\end{align*}

\begin{figure}[!ht]
\centering

\begin{tikzpicture}[xscale=1.5]
\node at (3,-1.5) { zeros };
\node at (1,-3.5) { non-zeros };
\node at (-0.2,-2.5) { $j$ };
\node at (2.2,0.2) { $j^*$ };

\draw (0,0) -- (0,-5);
\draw (0,-5) -- (4,-5);
\draw (4,-5) -- (4,0);
\draw (0,0) -- (4,0);
\draw (0,0) -- (0,-1);
\draw (0,-1) -- (1,-1);
\draw (1,-1) -- (1,-2);
\draw (1,-2) -- (2,-2);
\draw (2,-2) -- (2,-3);
\draw (2,-3) -- (3,-3);
\draw (3,-3) -- (3,-4);
\draw (3,-4) -- (4,-4);
\draw (4,-4) -- (4,-5);
\draw (0,-2.3) -- (4,-2.3);
\draw (0,-2.7) -- (4,-2.7);
\draw (2,0) -- (2,-5);
\draw (2.3,0) -- (2.3,-5);

\node at (2.15,-2.5) { $0$ };
\draw[decorate, decoration={brace, amplitude=1ex, raise=1ex}]
  (0, 0) -- (2, 0) node[pos=.5, above=2.5ex] { $supp(j)$ };
\end{tikzpicture}

\caption{The support $supp(j)$ and the critical value $j^*$ of a row $j$ in the vertical column arrangement of the step vectors $\bm{b}^1$,\dots,$\bm{b}^K$.}
\label{fig:stairswithsupport}
\end{figure}
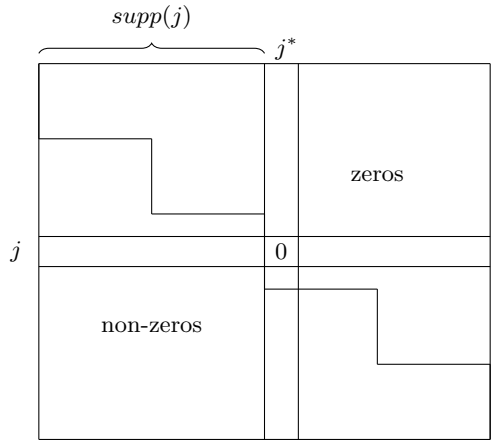

The proof is based on three observations:

1) We can assume that $Q$ contains only constraints of the form $x_k \geq x_j^n$ with $j < k$. If $Q$ contains a constraint $x_k \geq x_j^n$ with $j \ge k$ then we would have $x_j \ge x_k \ge x_j^n \ge x_j$ and thus $x_j = x_k = 1$ or $x_j = x_k = 0$. Thus, these can be guessed and substituted by the NP procedure.

\newcommand{\thei}{{l}}

2) If there is $x_j^n \le x_k \in Q$ such that $I = \text{supp}(j) = \text{supp}(k)$ then $\alpha_i \le m$ for every $i \in I$. Towards a contradiction, assume that $\alpha_{\thei} \geq m+1$ for some $\thei \in I$. 
Note that $\thei < k < j$ and $b^\thei_j > 0$.
Let $v_j = a_j + \alpha_{\thei} b_j^{\thei}$ and $v_k = a_k + \alpha_{\thei} b_k^{\thei}$. We have $v_j^n > v_k$ because: 
\[
v_j^n \geq \alpha_{\thei}^n > (\alpha_{\thei} - 1)(\alpha_{\thei}+1)\alpha_{\thei}^{n-2} \ge m(\alpha_{\thei}+1)\alpha_{\thei}^{n-2} \geq 
m(\alpha_{\thei}+1) = m + \alpha_{\thei} m \ge v_k
\]
It is also the case that $v_j \ge \alpha_l b_j^\thei \ge \alpha_l \ge m+1$.

Since $(x_j,x_k) = (v_j,v_k) + \sum_{i \in I \setminus \{\thei\}} \alpha_i (b^i_j,b^i_k)$, we obtain a contradiction with the inequality $x_j^n \le x_k$:
\begin{align*}
x_j^n 
&= 
\Big(v_j + \sum_{i \in I \setminus \{\thei\}} \alpha_i b_j^i\Big)^n \\ 
&= \Big(v_j + \sum_{i \in \text{supp(j)} \setminus \{\thei\}} \alpha_i b_j^i\Big)^n \\
&\ge v_j^n + \binom{n}{n-1} v_j^{n-1} \Bigg(\sum_{i \in \text{supp(j)} \setminus \{\thei\}} \alpha_i \Bigg) 
    + \Bigg( \sum_{i \in \text{supp(j)} \setminus \{\thei\}} \alpha_i \Bigg)^n \\ 
& > v_j^n + v_j \sum_{i \in \text{supp(j)} \setminus \{\thei\}} \alpha_i \\ 
&\ge  v_k + (m+1) \sum_{i \in \text{supp}(k) \setminus \{\thei\}} \alpha_i \\ 
&\ge v_k + \sum_{i \in \text{supp(k)} \setminus \{\thei\}} \alpha_i b_k^i \\& = x_k
\end{align*}

3) Otherwise, for every $x_j^n \le x_k \in Q$, $\text{supp}(j) \subsetneq \text{supp}(k)$. Then, $x_j$ depends only on  $b^1,\ldots,b^{j^*-1}$ while $x_k$ depends also on a term $\alpha_{j^*} b_k^{j^*}$ where $b_k^{j^*} > 0$. We can thus extend any solution $(\alpha_1,\ldots,\alpha_{j^*-1})$ of constraints which only depends on $b^1, \ldots, b^{j^*-1}$ to a solution $(\alpha_1,\ldots,\alpha_{j^*})$ where $x_j^n \le x_k$ also holds, by making $\alpha_{j^*}$ large enough.

These observations suggest the following NP algorithm.

\vspace{.5em}



On input $\langle L \land Q \rangle$ in normal form:

\begin{enumerate}
    \item Compute the set $B$ of inequalities $x_j^n \le x_k \in Q$ such that $supp(j) = supp(k)$.
    
    \item If $B = \emptyset$ then accept.
    Otherwise, non-deterministically guess $\alpha_1,\ldots,\alpha_l \leq m$ where $l = \max\limits_{x_j^n \le x_k \in B} (j^* - 1)$.
    
    \item Accept iff $\alpha_1,\ldots, \alpha_l$ satisfy the inequalities $x_j^n \le x_k \in Q$ with $k^* - 1 \le l$.
\end{enumerate}

If there is a solution to the constraints in $Q$ then it is clear that the algorithm accepts. Conversely, if the algorithm accepts, we can construct a solution $(\alpha_1,\ldots,\alpha_l, \alpha_{l+1}^*, \ldots, \alpha_n^*)$ for $Q$ as follows.

\vspace{.5em}
On input $\langle Q, B \rangle$:

\begin{enumerate}
    \item Sort the  inequalities $x_j^n \le x_k \in Q \setminus B$ with $k > l$ by lexicographic order of the tuple $(j,k)$ in a list $L$.

    \item If $L$ is empty set all remaining $\alpha_j^*$ to zero and stop. 
    \item Otherwise, remove the first element $(j,k)$ of $L$ and find a coefficient $\alpha_{jk}$ for $b^{j^*}$ such that $x_j^n \le x_k$ is satisfied. This is possible since $supp(j) \subset supp(k)$.
    \item Repeat for all pairs of the form $(j',k')$ with $supp(j') = supp(j)$. Since the step vectors are ordered lexicographically, these appear immediately after $(j,k)$.
    
    \item Set $\alpha_{j^*} = \max\limits_{j',k} \alpha_{j',k}$ and $\alpha_{t} = 0$ for any $l+1 \le t < j^*$ previously left unset.
    \item Go back to step 2.
\end{enumerate}

The result, $(\alpha_1,\ldots,\alpha_l, \alpha_{l+1}^*, \ldots, \alpha_n^*)$, satisfies $Q$ by construction.
\end{proof}

The constraints we solve in Theorem~\ref{thm:more} are of the form $x \ge y^n$ or equivalently $0 \ge f(x,y)$ where
$f(x,y) = y^n -x$. This function is convex since it is the addition of a linear function (trivially convex) and the $n^{\textit{th}}$ power function (convex for having a positive semidefinite Hessian) \cite[sections 3.1.4 and 3.2.1]{boyd_convex_2004}. 

In \cite{khachiyan_polynomial_1980}, a related result is given for systems of $s$ convex polynomial inequalities $f_i(x_1,\ldots,x_n) \le 0, i = 1, \ldots, s$ where the $f_i(x_1,\ldots,x_n) \in \mathbb{Z}[x_1,\ldots,x_n]$ are convex polynomials in $\mathbb{R}^n$ with integral coefficients. It is formulated over the integers but one can add the linear constraints $-x_i \le 0$ (which are trivially convex) to obtain an analogous result over the natural numbers.

\begin{theorem}[Tarasov and Khachiyan (1980) \cite{tarasov_bounds_1980}]
\label{thm:khachiyan}
For a fixed $d \ge 1$ the problem of determining the consistency of systems of convex diophantine inequalities of degree at most $d$ over the integers belongs to the class NP.
\end{theorem}

While Theorem~\ref{thm:khachiyan} allows for arbitrary convex inequalities, it fixes the degree that the polynomial-time verifier can handle. Our Theorem~\ref{thm:more}, on the other hand, focuses on monomial constraints but gives a single verifier for the entire class, over all degrees $d$.

\section{Satisfiability of Non-Convex Monomials}
\label{section:less}

This section proves an NP complexity bound for less-than-monomial constraints. Our proof shows a \textit{small model property} for $(\alpha_1,\ldots,\alpha_n)$. If there is a solution, there is a solution in the ball centered at the origin with radius of binary length $\log(m+1)$. The ball is taken with respect to the infinite norm $\|(\alpha_1,\ldots, \alpha_n)\|_{\infty}$.

The key insight of the proof is that we can avoid recomputation of the underlying linear set each time we substitute one fixed variable. Instead, we guess small coefficients $\alpha_i$ and show that if $\alpha_i$ is large enough then the prequadratic constraints $x_l \le x_j \cdot x_k$ where $b_l^i, b_j^i, b_k^i > 0$ are satisfiable. This follows from an inductive argument that is sketched in the fourth case distinction below.

\begin{theorem} \label{thm:less}
Satisfiability of less-than-monomial constraints is in NP.
\end{theorem}
\begin{proof}
We can assume that the input formula is of the form specified in Lemma~\ref{lem:normal}. Let $m$ denote the maximum of the coefficients of the generators of the linear part. We introduce the notation $i_*$ to refer to the row of the last zero entry and $null(\bm{b}^i)$ to refer to the set of indices with zero values of the step vector $\bm{b}^i$ (see figure \ref{fig:stairswithnullity}):

\begin{align*}
i_* &= 
\begin{cases}
0              & \text{ if } \text{null}(\bm{b}^i) = \emptyset \\
\max \text{null}(\bm{b}^i) & \text{ if } \text{null}(\bm{b}^i) \neq \emptyset 
\end{cases} \\
\text{null}(\bm{b}^i) &= \{j \mid \bm{b}_j^i = 0\} 
\end{align*}

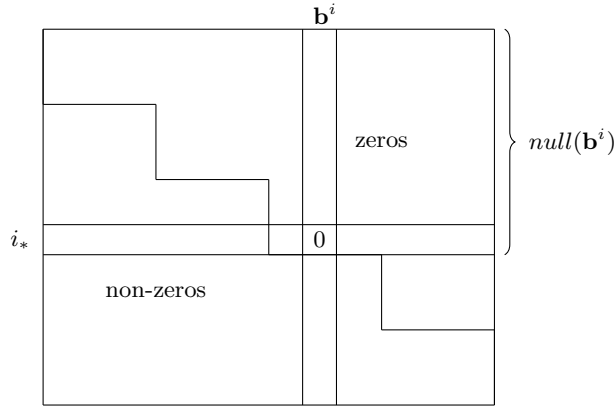
\begin{figure}[!ht]
\centering

\begin{tikzpicture}[xscale=1.5]
\node at (3,-1.5) { zeros };
\node at (1,-3.5) { non-zeros };

\node at (-0.2,-0.1) {  };
\node at (-0.2,-0.85) {  };
\node at (-0.2,-1.85) {  };
\node at (-0.2,-2.8) { $i_*$ };
\node at (-0.2,-3.85) {  };
\node at (-0.2,-4.85) {  };

\node at (0.15,0.2) {  };
\node at (0.85,0.2) { };
\node at (1.85,0.2) {  };
\node at (2.5,0.2) { $\bm{b}^i$ };

\node at (1.85,-2.85) {  };

\node at (2.45,-2.8) { $0$ };

\draw (0,0) -- (0,-5);
\draw (0,-5) -- (4,-5);
\draw (4,-5) -- (4,0);
\draw (0,0) -- (4,0);
\draw (0,0) -- (0,-1);
\draw (0,-1) -- (1,-1);
\draw (1,-1) -- (1,-2);
\draw (1,-2) -- (2,-2);
\draw (2,-2) -- (2,-3);
\draw (2,-3) -- (3,-3);
\draw (3,-3) -- (3,-4);
\draw (3,-4) -- (4,-4);
\draw (4,-4) -- (4,-5);
\draw (0,-2.6) -- (4,-2.6);
\draw (0,-3) -- (4,-3);
\draw (2.3,0) -- (2.3,-5);
\draw (2.6,0) -- (2.6,-5);

\draw[decorate, decoration={brace, amplitude=1ex, raise=1ex}]
  (4, 0) -- (4, -3) node[pos=.5, right=2.5ex] { $null(\bm{b}^i)$ };
\end{tikzpicture}

\caption{The nullity $null(\bm{b}^i)$ and the critical value $i_*$ of a column $\bm{b}^i$ in the vertical column arrangement of the step vectors $\bm{b}^1$,\dots,$\bm{b}^K$.}
\label{fig:stairswithnullity}
\end{figure}

Given a solution $\bm{x^s} = \bm{a} + \sum_{i = 1}^k \alpha_i \bm{b}^i$, our goal is to prove that
there exists another solution $\bm{{x^s}'} = \bm{a} + \sum_{i = 1}^k \alpha_i' \bm{b}^i$ of $L \land Q$ where
$\max\limits_i \alpha_i' \le m+1$.

\newcommand{\thei}{{l}}

If $\max\limits_i \alpha_i \le m+1$ then we are done. Otherwise, let $\thei$ be the smallest index such that $\alpha_{\thei} > m+1$. Since we assume a lexicographic order in the $\bm{b}^i$'s, if $i \le i'$ then $\text{null}(\bm{b}^{i}) \subseteq \text{null}(\bm{b}^{i'})$. This together with the linear order in the solutions $\bm{x^s}$ leads to a matrix of step vectors where $b_{\thei_*}^{l}$ acts as a pivotal element: it separates the lower-left non-zero submatrix from the upper-right zero part.


We construct another solution 
$\bm{x^{s'}} = \bm{a} + \sum_{i = 1}^k {\alpha_i}' \bm{b^i}$ with:
\[
\alpha_{i}' = \begin{cases}
\alpha_i & \text{ if } i < i_* \\
m+1 & \text{ if } i = i_* \\
 0 & \text{ if } i <i_* 
\end{cases}
\]

$\bm{x^{s'}}$ is a small solution in terms of $(\alpha_1,\ldots,\alpha_n)$ since $\| (\alpha_1',\ldots,\alpha_n') \|_{\infty} = m+1$. Furthermore, $x_j^{s'} \leq x_j^s$ for any $j$ since if $j \le \thei_*$ then:
\[
x_j^{s} = a_j + \sum_{i = 1}^k \alpha_i b_j^i = a_j + \sum_{i = 1}^{\thei_*-1} \alpha_i b_j^i = a_j + \sum_{i = 1}^{\thei_*-1} \alpha_i' b_j^i
= a_j + \sum_{i = 1}^k \alpha_i' b_j^i = x_j^{s'}
\]
and if $\thei_* < j$:

\begin{align*}
x_j^{s'} &= a_j + \sum_{i = 1}^k \alpha_i' b_j^i = a_j + \sum_{i < \thei} \alpha_i b_j^i + (m+1) b_j^{\thei} < \\ &< a_j + \sum_{i < \thei} \alpha_i b_j^i + \alpha_{\thei} b_j^{\thei} \le a_j+ \sum_{i = 1}^k \alpha_i b_j^i = x_j^s
\end{align*}
where we used that all base and step vector components and all coefficients are greater or equal than zero and $\alpha_{\thei} > m+1, b_j^i \ge 1$ for $i < \thei$.
    
Finally, we show that $\bm{x^{s'}}$ is a solution of $Q$. Given $x_{j} \leq x_{k}^n \in Q$, we show $x^s_j \leq ({x^s_k})^n$. Consider four cases:

\begin{enumerate}
    \item $j \le k$: the components of the solutions are linearly ordered and thus $x_j \le x_k \le x_k^n$.
    \item $k < j \le \thei_*$: $x_j^{s'} = x_j^s \le (x_k^s)^n = (x_k^{s'})^n$.
    \item $k \le \thei_* < j$: $x_j^{s'} < x_j^s \le (x_k^s)^n = (x_k^{s'})^n
    $
    \item $\thei_* < k < j$: call $v_j = a_j + (m+1)b^{\thei}_j$ and $v_k = a_k + (m+1)b^{\thei}_k$.
    
    We show by finite induction on the natural number $t \le i_*$ that: 
    \[
    v_j + \sum_{i < t} \alpha_i' b_j^i \le \Big(v_k + \sum_{i < t} \alpha_i' b_k^i\Big)^n
    \]

    \begin{enumerate}
        \item In the base case, $t = 0$ and we need to show $v_j \le v_k^n$: 
        \[
        v_j \le m + (m+1) m \le (m+1)^2 \le (m+1)^n \le (a_k + (m+1) b_k^{\thei})^n = v_k^n
        \]
        
        \item Assume that for $t < \thei$, we have: 
        \[
        v_j + \sum_{i < t} \alpha_i' b_j^i \le \Big(v_k + \sum_{i < t} \alpha_i' b_k^i\Big)^n
        \]
        then we need to show that:
        \[
        v_j + \sum_{i < t+1} \alpha_i' b_j^i \le \Big(v_k + \sum_{i < t+1} \alpha_i' b_k^i\Big)^n
        \]
        Set $v_j' = v_j + \sum_{i < t} \alpha_i' b_j^i$ and $v_k' = v_k + \sum_{i < t} \alpha_i' b_k^i$. Then it suffices to show that:
        
        \begin{align*}
        v_j' + \alpha_{t}' b_j^{t} &\le v_k'^n + \alpha_{t}' m \\ &\le v_k'^n + \alpha_{t}' v_k \\ &\le v_k'^n + \binom{n}{n-1} v_k'^{n-1} \alpha_{t}' b_k^{t} \le (v_k' + \alpha_{t}' b_k^{t})^n
        \end{align*}
        
         where in the second and third inequalities we have used that since $k > \thei_*$ and $t, i_* \le i_*$ we have that  $b_k^{\thei}, b_k^{t} \ge 1$.
    \end{enumerate}
    \end{enumerate}

Thus, the trivial NP procedure that guesses all the coefficients $\alpha_1,\ldots,\alpha_n \leq m + 1$  and accepts if and only if $\bm{x^s} = \bm{a} + \sum_{i = 1}^k \alpha_i \bm{b}^i$ respects the inequalities $x_j \le x_k^n \in Q$ shows the problem can be decided in NP.
\end{proof}

The function $f(x,y,z,\ldots) = y^n - x$ is convex as discussed in Section~\ref{section:more}. The constraints of the form $f(x,y,z,\ldots) \ge 0$ are called reverse convex in the operations research literature. To the best of our knowledge this is the first complexity result for conjunctions of reverse convex constraints over the integers.

\newcommand{\thei}{{l}}

Note that it is key that, thanks to the inductive argument, we can disregard the remaining $\alpha$'s after $\alpha_{\thei}$. These $\alpha$'s would be detrimental for an inequality $x_j \le x_k^n$ with $k < \thei_* < j$. However, in the general case, we could furthermore have linear inequalities $x_j \le x_k x_m$ with $m < \thei_* < k,j$ and we cannot guarantee that the $\alpha$'s after $\alpha_{\thei}$ are superfluous. Furthermore, the inductive argument would fail in the case that $b^i_j > 0$ but $b^i_k = b^i_m = 0$.

\section{Satisfiability of Monomial Inequalities with Solved Linear Constraints}
\label{section:hardness}

In previous sections, we have presented decision procedures that leveraged insights on the structure of the set of solutions of linear constraints in order to find solutions to restricted families of non-linear inequalities. It is thus natural to ask how hard it is to check satisfiability of the non-linear part when given the set of solutions to the linear constraints as input.

In this section, we show that the answer to this question is mixed. On the one hand, we observe that from the results of \cite{onn_nonlinear_2010}, it follows that for a single more-than-monomial constraint, satisfiability with the Hilbert basis given as input can be decided in polynomial time. On the other hand, we show that this is no longer true when given more-than-monomial,  less-than-monomial or general convex constraints. For each of these cases, we show that the resulting problem is NP-hard.

\subsection{One Monomial Inequality}

We start with the case where there is a single monomial inequality and the linear part has been solved in the normal form suggested, i.e. we have:
\[
\begin{cases}
x_k \ge x_j^l \\
\mathbf{x} = \mathbf{a} + \sum \alpha_i \mathbf{b^i} 
\end{cases}
\]
If $supp(j) \neq supp(k)$ then we know there is a solution. If $supp(j) = supp(k)$ then, by the second observation in the proof of Theorem~\ref{thm:more}, a solution necessarily lies in the ball $B(0, m+2nm^2 \log(4m))$. Theorem 3.12 in \cite{onn_nonlinear_2010} shows that in contrast to Sections~\ref{section:convexhard}, \ref{section:morehard} and \ref{section:lesshard}, this instance can be solved in polynomial time:

\begin{theorem}[Onn \cite{onn_nonlinear_2010}]
There is an algorithm that, given $A \in \mathbb{Z}^{m \times n}$, $\mathcal{G}(A), l, u \in \mathbb{Z}^{n}, b \in \mathbb{Z}^{m}$ and separable convex $f: \mathbb{Z}^{n} \rightarrow \mathbb{Z}$ presented by comparison oracle, solves in time polynomial in $\langle A, \mathcal{G}(A), l, u, b, \hat f \rangle$ the problem $
\min \left\{f(x): x \in \mathbb{Z}^{n}, A x=b, l \leq x \leq u\right\} .
$
\end{theorem}

Here $\mathcal{G}(A)$ stands for the so-called Graver basis which is a generalisation of the notion of Hilbert basis for the non-positive orthants. On the other hand, $\hat f$ stands for the maximum of $f$ over the compact domain $l \le x \le u$. The theorem guarantees that the minimisation problem can be solved in polynomial time in the size of the parameters. Since we are interested in the solution in a ball, the maximum of the function $\hat f$ is simply a constant and can be ignored. Then, we would minimise the function $f(x) = x_j^l - x_k$. If the minimum value is $\le 0$ then we accept, otherwise we reject. 

\subsection{General Convex Constraints} \label{section:convexhard}

When arbitrary non-linear constraints are included, NP-hardness is relatively easy to show depending on the richness of the range of convex functions we use. If one considers the class of convex quadratics as in \cite{l_g_khachiyan_convexity_1983} then one single quadratic suffices to show NP-hardness. We call this problem convex quadratic consistency (CQC):


\begin{definition}
CQC is the problem of deciding if there are integer solutions to the inequality:
\[
f(x_1,\ldots,x_n) \le 0
\]
for $f$ a convex quadratic.
\end{definition}

Recall the definition of subset sum from \cite{garey_computers_1990}:

\begin{definition}
SUBSETSUM is the decision problem which for a set of positive integers $X = \{x_1,\ldots, x_l\} \subseteq \mathbb{Z}^+$ and a bound $s \in \mathbb{Z}^+$ given as input determines whether there exists a subset $A \subseteq X$ such that
$\sum_{a \in A} a = s$.
\end{definition}


To show NP-hardness, we reduce to CQC the following variant of subset sum:

\begin{definition}
$\text{SUBSETSUM}_{\mathbb{Z}}$ is the decision problem which for a set of integers $X = \{x_1,\ldots,x_l\} \subseteq \mathbb{Z}$ given as input determines whether there exists a subset $A \subseteq X$ such that $\sum_{a \in A} a = 0$.
\end{definition}

The NP-hardness of this variant follows by reducing SUBSETSUM to $\text{SUBSETSUM}_{\mathbb{Z}}$, mapping each set $X \subseteq \mathbb{Z}^+$ and bound $s \in \mathbb{Z}^+$ to the set $X \cup \{-s\} \subseteq \mathbb{Z}$. 

\begin{lemma}
$\text{SUBSETSUM}_{\mathbb{Z}}$ is NP-hard.
\label{lem:subsethard}
\end{lemma}

Finally, we reduce $\text{SUBSETSUM}_{\mathbb{Z}}$ to CQC:

\begin{proposition}
CQC is NP-hard.
\end{proposition}
\begin{proof}
We reduce $\text{SUBSETSUM}_{\mathbb{Z}}$ to $\text{CQC}$. For this, we map each subset $X = \{x_1,\ldots,x_l\} \subseteq \mathbb{Z}$ to the quadratic inequality:
\begin{equation} \label{eq:1}
\sum_{i = 1}^l (x_i^2-x_i) + \Big(\sum_{i = 1}^l a_i x_i\Big)^2 \le 0
\end{equation}

Observe that the left hand side of the inequality (\ref{eq:1}) is a convex quadratic. In particular, it is convex since it is a sum of two convex functions: the first addend is a sum of convex functions and the second addend is the composition of a convex function with an affine transformation \cite[section 3.3.2]{boyd_convex_2004}. 

If there is a subset $A \subseteq X$ such that $\sum_{a \in A} a = 0$ then we can find $x_i \in \{0,1\}$ such that $\sum_{i = 1}^l a_i x_i = 0$ and thus the left hand side of the inequality (\ref{eq:1}) is equal to $0$. Conversely, if inequality (\ref{eq:1}) has a solution then necessarily $x_i \in \{0,1\}$ and $\sum_{i=1}^l a_i x_i = 0$. This gives a solution to the $\text{SUBSETSUM}_{\mathbb{Z}}$ problem.
\end{proof}

\subsection{More-Than-Monomial Constraints} \label{section:morehard}

We cannot do a similar reduction in the more-than-monomial problem since in that case, all the $a_i$'s need to be non-negative. But we can still show NP-hardness. Indeed,  consider a family of more-than-monomial constraints:
\[
\begin{cases}
\{x_k \ge x_j^{n_i}\}_{i = 1, \ldots q, n_i \in \mathbb{N}, n_i \ge 2} \\
\mathbf{x} = \mathbf{a} + \sum \alpha_i \mathbf{b^i} 
\end{cases}
\]

To show NP-hardness we reduce from the circuit satisfiability problem \cite{arora_computational_2009}:

\begin{definition}
CKT-SAT is the decision problem which for a given $n$-input circuit $C$ determines whether there exists $u \in \{0,1\}^n$ such that $C(u) = 1$.
\end{definition}



\begin{theorem}
More-than-monomial is NP-hard.
\end{theorem}
\begin{proof}
We reduce CKT-SAT to more-than-monomial. In order to ease the translation, we assume that the circuit to which the reduction is applied is given in terms of NAND gates. It is known that NAND gates are universal, that is, any circuit can be represented in terms of this operation. Since translating each basic gate requires only a constant number of NAND gates, one further observes that the translation of a Boolean circuit into an equivalent NAND-based circuit increases size by a constant multiplicative factor, which is irrelevant for complexity considerations.

First, we observe that we can encode each NAND gate with polynomially many more-than-monomial constraints. 

Let $g: z = \lnot (x \land y)$ be a NAND gate. We introduce four variables $\alpha_0, \alpha_1, \alpha_2, \alpha_3$. The index $i$ of $\alpha_i$ translated to a two-digit binary number corresponds to each possible valuation of $x,y$. We add the equalities $x = \alpha_2 + \alpha_3, y = \alpha_1 + \alpha_3, z = \alpha_1 + \alpha_2 + \alpha_3$.

We impose for each $i,j \in \{0,1,2,3\}$ ($i \neq j$) the restriction that $\alpha_i + \alpha_j \le 1$. This ensures that at most one coefficient $\alpha_i$ is set to one. This restriction can be enforced with more-than-monomial constraints by adding variables $u_{ij},v_{ij}$ with $i \neq j$ such that $u_{ij} = \alpha_i + \alpha_j, v = 3, u_{ij}^2 \le v_{ij}$.

Similarly, we impose the restriction that $1 \le \alpha_0 + \alpha_1 + \alpha_2 + \alpha_3$. This ensures that at least one coefficient is satisfied. This restriction can be enforced by adding variables $r,s$ such that $r = 1, s = \alpha_0 + \alpha_1 + \alpha_2 + \alpha_3, r^2 \le s$.

\begin{equation} \label{eq:linear}
\begin{pmatrix}
x \\
y \\
z \\
u_{01} \\
\vdots \\
u_{32} \\
v \\
r \\
s
\end{pmatrix} = 
\begin{pmatrix}
0 \\
0 \\
0 \\
0 \\
\vdots \\
0 \\
3 \\
1 \\
0
\end{pmatrix}
+
\alpha_0
\begin{pmatrix}
0 \\
0 \\
0 \\
1 \\
\vdots \\
0 \\
0 \\
0 \\
1
\end{pmatrix}
+
\alpha_1
\begin{pmatrix}
0 \\
1 \\
1 \\
1 \\
\vdots \\
0 \\
0 \\
0 \\
1
\end{pmatrix}
+
\alpha_2
\begin{pmatrix}
1 \\
0 \\
1 \\
0 \\
\vdots \\
1 \\
0 \\
0 \\
1
\end{pmatrix}
+
\alpha_3
\begin{pmatrix}
1 \\
1 \\
1 \\
0 \\
\vdots \\
1 \\
0 \\
0 \\
1
\end{pmatrix}
\end{equation}

In summary, the linear set of equation~\ref{eq:linear} together with the prequadratic constraints $r^2 \le s, u_{ij}^2 \le v$ where $i < j$ and $i,j \in \{0,\ldots,3\}$
encode the operation of $g$.

Second, we encode the rest of the circuit. For each new gate, we add a new diagonal block to the step vectors. Each block repeats the pattern shown in equation~\ref{eq:linear}.

We may reuse any of the variables $x,y,z$ in other gates. To do so, we need to encode equality between two variables of the left hand side. Since we will later enforce that each variable is zero-one valued, this can be done using more-than-monomial constraints: to say that $x$ and $y$ are equal it suffices to impose that $x^2 \le y$ and $y^2 \le x$. In the zero-one valued case, this implies that $x = y$.

The last step of the transformation ensures that all variables, either those labelling wires in the original circuit or those added later, are zero-one valued. In particular, for the coefficients $\alpha_i$ of the linear set, we first introduce equations $t = \alpha_i$. Finally, we add the inequalities $x_i^2 \le x_i$ for all the resulting variables.

The transformation can be clearly done in polynomial time and the correctness is ensured by construction. Thus, more-than-monomial is NP-hard even when the underlying linear set is explicitly given, as we wanted to show.
\end{proof}

\subsection{Less-Than-Monomial Constraints} \label{section:lesshard}

Now assume that we are given a family of monomials:
\[
\begin{cases}
\{x_j \le x_k^{n_i}\}_{i = 1, \ldots q, n_i \in \mathbb{N}, n_i \ge 2} \\
\mathbf{x} = \mathbf{a} + \sum g_i \mathbf{b^i} 
\end{cases}
\]

\begin{theorem}
Less-than-monomial is NP-hard.
\end{theorem}
\begin{proof}
It suffices to modify slightly the construction above. To enforce $x_j \in \{0,1\}$, it suffices to set $x_i = 1$ and $x_j \le x_i^2$. To enforce $\alpha_i + \alpha_j \le 1$ it suffices to write $u_{ij} = \alpha_i + \alpha_j, v = 1, u_{ij} \le v^2$. To enforce that $1 \le \alpha_0 + \alpha_1 + \alpha_3 + \alpha_3$ we simply set $r = 1, s = \alpha_0 + \alpha_1 + \alpha_2 + \alpha_3, r \le s^2$.
\end{proof}

\section{Logical Consequences}
\label{section:logical}

Theorems~\ref{thm:more} and \ref{thm:less} can be used to establish an NP complexity bound on some fragments of theories of relational logic since an unconstrained $n$-ary relation $R$ on a set $\mathcal{U}$ exists if and only if $|R| \le |\mathcal{U}|^n$. Let's consider as in \cite{yessenov_collections_2010}, the theory of QFBAPA enriched with unary functions of sets and their inverse and direct function images $f^{-1}[B] = \{ y | \exists x. x \in B \land y = f(x) \}$ and
$f[B] = \{ y | \exists x. x \in B \land y = f(x) \}$. Let's also allow for a set variable $B$, to form the Cartesian product $B^n$ of $B$ iterated $n$ times. Then the satisfiability of the formula $S = f^{-1}[P^n]$ is equivalent to the satisfiability of the non-linear constraint $|P|^n \le |S|$. Similarly, the satisfiability of the formula $S = f[P^n]$ is equivalent to the satisfiability of the non-linear constraint $|S| \le |P|^n$.

As a result, we obtain a fragment that is strictly more expressive than the language of QFBAPA. It enriches the language of Figure~\ref{fig:qfbapa-syntax} with top-level constraints of the form $S = f^{-1}[P^n]$ or $S = f[P^n]$. Note that one cannot mix both kinds of constraints since as remarked in the introduction this would express Hilbert's 10th problem and would thus yield an undecidable fragment.





\section{Conclusion}
\label{section:conclusion}

Non-linear Diophantine constraints have been widely investigated in mathematical optimisation and automated reasoning. Despite the number of applications of prequadratic \cite{givan_tarskian_2002, arenas_complexity_2008, david_efficient_2012 ,yessenov_collections_2010, raya_np_2022} and more general constraints \cite{gurari_np-complete_1978, l_g_khachiyan_convexity_1983, baeten_solvable_2003, bugliesi_intersection_2006,  onn_nonlinear_2010} there exist few classes in the literature with low complexity bounds making them suitable for integration in satisfiability modulo theory solvers \cite{bradley_calculus_2007, hutchison_z3_2008, shoukry_smc_2017, cimatti_incremental_2018, borralleras_incomplete_2019, barbosa_cvc5_2022}. In this work, we prove an optimal bound for a subfamily of prequadratic Diophantine constraints. We show that these constraints are useful in analyzing the cardinality of cartesian powers, which can be used in fragments of Boolean algebra with function and inverse images. We have remarked that in the case of a single monomial constraint, the complexity is polynomial when given the Hilbert basis of the linear part. On the other hand, we have shown that with arbitrary monomial constraints the problem becomes NP-hard even if the Hilbert basis of the linear part is given. The key of our development is the normal form of Section~\ref{section:normal}. In future work, we plan to investigate larger classes of (non-)convex and general prequadratic constraints.

\bibliography{refmore}

\end{document}